\documentstyle[floats,multicol,aps,epsf,prd]{revtex}

\def\ID#1{1\kern-.225em{\rm l}_{#1}}
\def\Tr{\mathop{\rm Tr}\nolimits}
\def\det{\mathop{\rm det}\nolimits}
\def\diag{\mathop{\rm diag}\nolimits}
\def\partialslash{\partial\kern-.5em\raise.2ex\hbox{/}}
\newfam\msbfam
\font\tenmsb=msbm10             \textfont\msbfam=\tenmsb
\font\sevenmsb=msbm7            \scriptfont\msbfam=\sevenmsb
\font\fivemsb=msbm5             \scriptscriptfont\msbfam=\fivemsb
\def\Bbb{\fam\msbfam \tenmsb}
\begin{document}
\draft
\twocolumn[\hsize\textwidth\columnwidth\hsize\csname
@twocolumnfalse\endcsname
\preprint{UTEXAS-HEP-99-11}
\title{Fermion masses in noncommutative geometry}
\author{R. Schelp}
\address{Department of Physics, University of Texas at Austin, Austin,
Texas 78712, USA}
\address{E-mail: schelp@physics.utexas.edu}
\date{\today}
\maketitle
\bigskip
\hrule
\begin{abstract}
{Recent indications of neutrino oscillations raise the question of the
possibility of incorporating massive neutrinos in the formulation of the
Standard Model (SM) within noncommutative geometry (NCG).  We find that
the NCG requirement of Poincar\'e duality constrains the numbers of 
massless quarks and neutrinos to be unequal unless new fermions are
introduced.  Possible scenarios in which this constraint is satisfied are 
discussed.}
\end{abstract}
\pacs{Keywords: Noncommutative geometry; Neutrino; Mass; Poincar\'e
duality}
\hrule
\bigskip
]
\section{Introduction}
The results of recent solar\cite{Kirsten99}, atmospheric\cite{Shiozawa99},
and accelerator-based\cite{Eitel00} neutrino experiments all suggest that
neutrino oscillations do occur. This leads one to the idea that, as
in the case for quarks, the neutrino flavor eigenstates are coherent
linear superpositions of mass eigenstates.  For a neutrino of flavor
$\ell$ we have
\begin{equation}
|\nu_\ell\rangle =\sum_{i=1}^NU_{\ell i}|\nu_i\rangle\;,
\end{equation}
where $U_{\ell i}$ is the leptonic mixing matrix.  A straightforward
quantum mechanical calculation leads to the following expression
\cite{BoehmVogel92} for the probablility that a neutrino of flavor $\ell$
with a momentum $p_\nu$ will oscillate into a neutrino of flavor $\ell'$
in a distance $L$:
\begin{eqnarray}
P_{\ell\ell'}(p_\nu ,L)=&&\sum_{i=1}^N|U_{\ell
i}U_{i\ell'}^\dagger|^2\\
+&&{\rm Re}\Biggl\lbrace\sum_{\scriptstyle i,j=1\atop\scriptstyle j\ne
i}^NU_{\ell i}U_{i\ell'}^\dagger U_{\ell j}U_{j\ell'}^\dagger
e^{i{|m_i^2-m_j^2|L\over 2p_\nu}}\Biggr\rbrace\;,\nonumber
\end{eqnarray}
where $m_i$ is the mass of the $i$th mass eigenstate.

In the usual SM, neutrinos are taken to be massless; experiments have only
given upper bounds on the masses.  However, nothing forces neutrinos to
be massless, and if one takes the oscillation experiments seriously, then
the analysis above shows that neutrinos must have mass.  The simplest way
to modify the SM to accomodate massive neutrinos is by including $\nu_R$
states and Yukawa couplings for the neutrinos (which would involve
the leptonic mixing matrix $U_{\ell i}$ above).

It has been shown that, while a NCG version of the SM is possible,
there are many Yang-Mills-Higgs models which cannot be formulated within 
the NCG framework\cite{Iochum96}.  In this article we address the question
of the extent to which the NCG formulation of the SM allows for massive
neutrinos.  This question has been addressed for an earlier NCG version of
the SM \cite{Gracia-Bondia95}.  Here we reconsider the issue in the
context of the requirement of Poincar\'e duality.

\section{Noncommutative Geometry}
We begin with a brief overview of the main ideas of NCG \'a la Connes as
they pertain to models of particle physics. We refer to \cite{Martin98}
for a clear and thorough review of NCG as it is applied to the SM.

The basic data required to specify a NCG are grouped together in a package
called a real spectral triple\cite{Connes95}.  A {\it spectral triple\/}
$({\cal A}, {\cal H}, D)$ consists of a Hilbert space ${\cal H}$ on which
the involutive algebra ${\cal A}$ is represented as a subalgebra of ${\cal
B}({\cal H})$, together with a self-adjoint operator $D$.  The operator
$D$ need not be bounded, but we require that $[D,a]$ be bounded for all
$a\in{\cal A}$ and that the resolvent $(D-\lambda)^{-1}$,
$\lambda\notin{\Bbb R}$ of $D$ is compact.  If there exists an operator
$\gamma$ on ${\cal H}$ which satisfies $\gamma^*=\gamma$, $\gamma^2=1$,
$D\gamma =-\gamma D$, and $\gamma a=a\gamma$ for all $a\in{\cal A}$, then
we classify $({\cal A}, {\cal H}, D)$ as a {\it even\/} spectral triple.
A {\it real\/} spectral triple is an even spectral triple for which there
also exists an antilinear isometry $J$ which satisfies $JD=DJ$,
$J^2=\epsilon$, $J\gamma=\epsilon'\gamma J$, where $\epsilon$ and
$\epsilon'$ take values in $\{+1,-1\}$ depending on the dimension of the
spectral triple. We also require that both $[a,JbJ^*]$ and $[[D,a],JbJ^*]$
vanish for all $a,b\in{\cal A}$.

The NCG particle physics models which have been constructed thus far all
make use of the notion of the product of real spectral triples.  One
factor of the product encodes the spacetime of the model and the other
factor encodes the internal space (the gauge group).  For the standard
model the components of the spectral triple are
\begin{eqnarray}
{\cal A}=&&{\cal A}_{\rm S}\otimes{\cal A}_{\rm F}=C^\infty({\Bbb
R}^4,{\Bbb R})\otimes({\Bbb C}\oplus{\Bbb H}\oplus M_3({\Bbb C}))\;,\\
{\cal H}=&&L^2({\Bbb R}^4,S)\otimes{\cal H}_{\rm F}\;,\\
D=&&\partialslash\otimes 1+\gamma_5\otimes D_{\rm F}\;,\\
\gamma=&&\gamma_5\otimes\gamma_F\;,\\
J=&&C\otimes J_F\;.
\end{eqnarray}
where ${\cal H}_{\rm F}$ is a complex Hilbert space which has a basis
labeled by the elementary fermions (including antiparticles), $D_{\rm
F}$ contains the Yukawa couplings, $\gamma_F$ is the grading by chirality
on ${\cal H}_{\rm F}$, $C$ is the charge conjugation operator on spinors,
and $J_{\rm F}$ is charge conjugation on ${\cal H}_{\rm F}$.

\section{Poincar\'e Duality}
In classical differential geometry, Poincar\'e duality is the requirement
that for $p$-forms $\alpha$ and $\beta$ the scalar product
\begin{equation}
(\alpha|\beta)=\int\alpha\wedge *\beta
\end{equation}
vanishes for all $\beta$ only if $\alpha=0$.  Connes has put forth seven
axioms defining NCGs \cite{Connes96} of which Riemannian geometries are
then seen as special cases.  The application of Connes' generalization of
the Poincar\'e duality condition to zero-dimensional spectral triples
(ones based on matrix algebras) is nicely explained in \cite{Martin98}.
Here we generalize the NCG formulation of the SM to allow for both
right-handed neutrinos and charge-2/3 quarks without right-handed states
and derive the implications of Poincar\'e duality in this setting.

Our first step is to adjust the representation of the algebra ${\cal
A}_{\rm F}={\Bbb C}\oplus{\Bbb H}\oplus M_3({\Bbb C})$.  Denote an element
of ${\cal A}_{\rm F}$ by the triple $(\lambda,q,m)$, where
$\lambda\in{\Bbb C}$, $q\in{\Bbb H}$, and $m\in M_3({\Bbb C})$.  Recall
also that a quaternion may be expressed as $q=\alpha +\beta j$, where
$\alpha,\beta\in{\Bbb C}$.  For $N_1$ generations of leptons without
right-handed neutrinos, the representation on the particle space is
\begin{equation}
\pi_{\ell 1}^+(\lambda,q)=
\bordermatrix{	   &\ell_R&\nu_R&\ell_L&\nu_L\cr
		\ell_R&\lambda&&&\cr
		\nu_R&&0&&\cr
		\ell_L&&&\alpha&\beta\cr
		\nu_L&&&-\bar\beta&\bar\alpha\cr}
	\otimes\ID{N_1}\;.
\end{equation}
In the remaining $N-N_1$ generations we include right-handed neutrinos,
thus the appropriate representation is
\begin{equation}
\pi_{\ell 2}^+(\lambda,q)=
\bordermatrix{     &\ell_R&\nu_R&\ell_L&\nu_L\cr
                \ell_R&\lambda&&&\cr
                \nu_R&&\bar\lambda&&\cr
                \ell_L&&&\alpha&\beta\cr
                \nu_L&&&-\bar\beta&\bar\alpha\cr}
        \otimes\ID{N-N_1}\;.
\end{equation}
In the above, $\ell$ denotes $e$, $\mu$, or $\tau$, not the weak doublets.
For the quarks, the representations are the same except that we replace
$N_1$ by $N_2$, the number of generations for which the charge-$2/3$ quark
has no right-handed state, and tensor each of the above by $\ID{3}$ for
color. We must also give the representations of $(\lambda,q,m)$ on the
antiparticles, which are simply multiplication by $\bar\lambda$ for the
leptons and multiplication by $m$ for the quarks.

The total Hilbert space of the theory is 
\begin{equation}
{\cal H}={\cal H}_\ell^+\oplus{\cal H}_q^+\oplus{\cal H}_\ell^-\oplus{\cal
H}_q^-
\end{equation}
where
\begin{eqnarray}
{\cal H}_\ell^+=&&{\Bbb C}_R\otimes{\Bbb C}^{N_1}\oplus{\Bbb
C}_L^2\otimes{\Bbb C}^{N_1}\\
	\oplus&&\,({\Bbb C}\oplus{\Bbb C})_R\otimes{\Bbb
C}^{N-N_1}\oplus{\Bbb C}_L^2\otimes{\Bbb C}^{N-N_1}\;\nonumber
\end{eqnarray}
for the leptons and
\begin{eqnarray}
{\cal H}_q^+=&&{\Bbb C}_R\otimes{\Bbb C}^{N_2}\otimes{\Bbb
C}_{\rm col}^3\oplus{\Bbb C}_L^2\otimes{\Bbb C}^{N_2}\otimes{\Bbb
C}_{\rm col}^3\\
        \oplus&&\,({\Bbb C}\oplus{\Bbb C})_R\otimes{\Bbb
C}^{N-N_2}\otimes{\Bbb C}_{\rm col}^3\oplus{\Bbb C}_L^2\otimes{\Bbb
C}^{N-N_2}\otimes{\Bbb C}_{\rm col}^3\nonumber
\end{eqnarray}
for the quarks. ${\cal H}_\ell^-$ and ${\cal H}_q^-$ are the
corresponding antiparticle Hilbert spaces.

The Poincar\'e duality condition is rooted in $K$-theory.  For a
zero-dimensional spectral triple the condition amounts to the
nondegeneracy of the intersection form
\begin{equation}
Q_{ij}=(p_i,p_j)=\Tr (\gamma p_iJp_jJ^*)\;,
\end{equation}
where $\gamma$ is the chirality operator, $J$, the charge conjugation 
operator, is the real structure on the spectral triple, and the
$p_i\,$s are generators of $K_0({\cal A})$.  For the finite part
of the algebra we are using, $K_0({\cal A}_{\rm F})={\Bbb Z}\oplus{\Bbb
Z}\oplus{\Bbb Z}$.  This group is generated by the minimal-rank
projections $1$ for ${\Bbb C}$, $\ID{2}$ for ${\Bbb H}$, and
$e=\diag(1,0,0)$ for $M_3({\Bbb C})$.

For our calculations we choose a basis in which $p_1=(-1)\oplus e$,
$p_2=1\oplus\ID{2}$, and $p_3=1$.  In terms $N_1$ and $N_2$ defined above,
the chirality and projections take the form
\begin{eqnarray}
\gamma\mapsto&&(1,0,-1,-1)^{N_1}\oplus (1,1,-1,-1)^{N-N_1}\nonumber\\
\oplus &&(1,0,-1,-1)^{3N_2}\oplus (1,1,-1,-1)^{3(N-N_2)}\\
\oplus &&(1,0,-1,-1)^{N_1}\oplus (1,1,-1,-1)^{N-N_1}\nonumber\\
\oplus &&(1,0,-1,-1)^{3N_2}\oplus (1,1,-1,-1)^{3(N-N_2)}\;,\nonumber\\
\nonumber\\
p_1\mapsto&&(-1,0,0,0)^{N_1}\oplus (-1,-1,0,0)^{N-N_1}\nonumber\\
\oplus &&(-1,0,0,0)^{3N_2}\oplus (-1,-1,0,0)^{3(N-N_2)}\\
\oplus &&(-1,0,-1,-1)^{N_1}\oplus (-1,-1,-1,-1)^{N-N_1}\nonumber\\
\oplus &&(e,0_3,e,e)^{N_2}\oplus (e,e,e,e)^{N-N_2}\;,\nonumber\\
\nonumber\\
p_2\mapsto&&(1,0,1,1)^{N_1}\oplus (1,1,1,1)^{N-N_1}\nonumber\\
\oplus &&(1,0,1,1)^{3N_2}\oplus (1,1,1,1)^{3(N-N_2)}\\
\oplus &&(1,0,1,1)^{N_1}\oplus (1,1,1,1)^{N-N_1}\nonumber\\
\oplus &&(0,0,0,0)^{3N_2}\oplus (0,0,0,0)^{3(N-N_2)}\;,\nonumber\\
\nonumber\\
p_3\mapsto&&(1,0,0,0)^{N_1}\oplus (1,1,0,0)^{N-N_1}\nonumber\\
\oplus &&(1,0,0,0)^{3N_2}\oplus (1,1,0,0)^{3(N-N_2)}\\
\oplus &&(1,0,1,1)^{N_1}\oplus (1,1,1,1)^{N-N_1}\nonumber\\
\oplus &&(0,0,0,0)^{3N_2}\oplus (0,0,0,0)^{3(N-N_2)}\;.\nonumber
\end{eqnarray}
These projections can be read off from the representations given above,
taking $(\lambda,q,m)$ to be $(-1,0,e)$ for $p_1$, $(1,\ID{2},0)$ for
$p_2$, and $(1,0,0)$ for $p_3$.

In order to calculate $Q$ we note that $J_{\rm F}$ acts by
\begin{equation}
J_{\rm F}\pmatrix{\xi\cr \bar\eta}=\pmatrix{\eta\cr\bar\xi}\;,
\end{equation}
where $(\xi,\bar\eta)\in{\cal H}^+\oplus{\cal H}^-$.  The effect on $p_i$
of conjugation by $J_{\rm F}$ is therefore to interchange the first four
terms in the direct sum with the last four terms.

Because the application of NCG methods to particle physics is relatively
new, we include here the details of the calculation of $Q_{12}$.  We work
out the argument of the trace in two pieces.  The first factor is
\begin{eqnarray}
\gamma p_1\mapsto&&(-1,0,0,0)^{N_1}\oplus (-1,-1,0,0)^{N-N_1}\nonumber\\
\oplus &&(-1,0,0,0)^{3N_2}\oplus (-1,-1,0,0)^{3(N-N_2)}\\
\oplus &&(-1,0,1,1)^{N_1}\oplus (-1,-1,1,1)^{N-N_1}\nonumber\\
\oplus &&(e,0_3,-e,-e)^{N_2}\oplus (e,e,-e,-e)^{N-N_2}\nonumber
\end{eqnarray}
and the second factor is
\begin{eqnarray}
Jp_2J^*\mapsto&&(1,0,1,1)^{N_1}\oplus (1,1,1,1)^{N-N_1}\nonumber\\
\oplus &&(0,0,0,0)^{3N_2}\oplus (0,0,0,0)^{3(N-N_2)}\\
\oplus &&(1,0,1,1)^{N_1}\oplus (1,1,1,1)^{N-N_1}\nonumber\\
\oplus &&(1,0,1,1)^{3N_2}\oplus (1,1,1,1)^{3(N-N_2)}\;.\nonumber
\end{eqnarray}
The product of these two factors is
\begin{eqnarray}
\gamma p_1Jp_2J^*
	&&\mapsto(-1,0,0,0)^{N_1}\oplus (-1,-1,0,0)^{N-N_1}\nonumber\\
	&&\oplus(0,0,0,0)^{3N_2}\oplus (0,0,0,0)^{3(N-N_2)}\\
	&&\oplus(-1,0,1,1)^{N_1}\oplus (-1,-1,1,1)^{N-N_1}\nonumber\\
	&&\oplus(e,0,-e,-e)^{N_2}\oplus (e,e,-e,-e)^{N-N_2}\;.
\end{eqnarray}
The trace of this product is
\begin{eqnarray}
\Tr (\gamma p_1Jp_2J^*)
	&&=-N_1-2(N-N_1)+N_1-N_2\\
	&&=-2N+2N_1-N_2\;.\nonumber
\end{eqnarray}
Performing similar calculations for the remaining elements, we find the
intersection form $Q$, omitting an overall factor of $-2$, to be
\begin{equation}
\pmatrix{
	N_1-N_2			& N-N_1+{1\over 2}N_2	
		& N-N_1+{1\over 2}N_2\cr
	N-N_1+{1\over 2}N_2	& N_1			& N_1-N\cr
	N-N_1+{1\over 2}N_2	& N_1-N			& N_1-2N\cr}\;.
\end{equation}

First of all, we note that for $N_1=N$ neutrinos without right-handed
states and $N_2=0$ quarks lacking right-handed states, as is the case in
the SM, we get
\begin{equation}
Q=2N\pmatrix{	-1 & 0  & 0\cr
		0  & -1 & 0\cr
		0  & 0  & 1\cr}\;,
\end{equation}
which agrees with the result of\cite{Connes95}.  The determinant of this
matrix is $8N^3$, so it is invertible and Poincar\'e duality condition is
satisfied.

If we keep right-handed states for all quarks (again, $N_2=0$) and let
$N_1=0$ so as to build an extension of the SM in which all neutrinos are
massive, the intersection form is
\begin{equation}
Q=2N\pmatrix{	0	& -1	& -1\cr
		-1	& 0	& 1\cr
		-1	& 1	& 2\cr}\;,
\end{equation}
recovering the result of Testard, as reported in \cite{Martin98}.  Here 
the determinant of $Q$ vanishes, indicating that the Poincar\'e duality
condition is {\it not\/} satisfied in this case.

In the general case we find that $\det{Q}=8(N_1-N_2)N^2$.  The requirement
that $Q$ be nondegenerate necessitates that the determinant not vanish.
Thus, we must have $N_1\ne N_2$.

\section{Possible Solutions}
There are three ways to satisfy this constraint while incorporating
massive neutrinos: (1) Keep all of the right-handed quark states ($N_2=0$)
and let one of the neutrinos (presumably $\nu_e$) remain massless 
($N_1=1$); (2) Give right-handed states to all of the neutrinos ($N_1=0$)
and let the up quark remain massless (so that $N_2=1$); (3) Give masses to
all quarks and leptons ($N_1=N_2=0$) and introduce new fermions (perhaps 
a sterile neutrino).  We now explore these possibilities.

{\it 1. No massless quarks, one massless neutrino\/}.  The conventional
wisdom with regard to neutrino oscillations is that neutrino masses are
necessary in order for them to occur.  In the standard analysis given
in the introduction, though, only mass differences play a role, allowing
for the possibility of oscillations between massless and massive
neutrinos.  Many more sophisticated treatments of neutrino oscillations
have been carried out (see the references in \cite{Kayser98}), but it
seems that in general they do not rely on neutrino masses, but mass
differences to generate oscillations.

With this in mind, we consider a case in which the eigenvalue of the first
of the three mass eigenstates is zero.  In this case there are three
mass-squared differences, $m_{12}^2$, $m_{23}^2$, and $m_{31}^2$, which
must sum to zero (here $m_{ij}^2=m_i^2-m_j^2$, where $m_i$ is the
eigenvalue of the $i$th mass eigenstate), leaving two independent masses.
Setting $m_1=0$ fixes the overall scale of the masses, but still leaves
two independent parameters.  By adjusting $m_2$ and $m_3$ we can account
for two different oscillation lengths; just as is the case were we to add
neutrino masses to the SM apart from the NCG context.  With just two mass
scales, we cannot explain the results of all three types of neutrino
oscillation experiments mentioned in the introduction, but we reserve
judgement on this possibility until more experimental verification is
available.

{\it 2. One massless quark, no massless neutrinos\/}.  Our analysis has
shown that Poincar\'e duality requires that $N_1\ne N_2$, so we may take
all neutrinos to be massive if we are willing to take $m_u=0$ (in
disagreement with the claim of \cite{Martin98} that the $m_u=0$
possibility is excluded).  There has been quite a bit of discussion in the
literature regarding the mass of the $u$-quark and the possibility of it
vanishing (e.g. \cite{Kaplan86,Donoghue92,Banks94}).  From the
standpoint of the strong CP problem, this option is quite attractive.  It
would lead to a solution in which the determinant of the quark mass matrix
vanishes\cite{Peccei89,Cheng88}.  It seems that there is currently no
phenomenological result which necessarily excludes the possibility of
$m_u=0$.

For us, though, the masslessness of the $u$-quark comes from the absence
of the right-handed component, not simply from the vanishing of the
appropriate Yukawa coupling.  This poses a problem for strong interation 
phenomenology, since it forces the couplings to right-handed and
left-handed quarks to be different and would thus lead to parity
violation in the strong interactions.

{\it 3. New fermions\/}.  If we take all of the neutrino experiments
seriously, it seems that not even three massive neutrinos are enough to
explain via oscillations all of the observed effects\cite{Zuber98}.  One
possible solution is then to include a new neutrino in the model.  This
option is severely limited by the experimental measurement of the $Z$
width, which is consistent with $2.993\pm 0.011$ light ($2m_\nu<m_Z$)
neutrino types\cite{Karlen98}.  We can avoid conflict with this result,
though, by introducing a neutrino which does not couple in the usual way
to the weak gauge bosons---a type of `sterile' neutrino, as was introduced
in\cite{Caldwell93,Peltoniemi93}. The effects of such an inclusion would
depend on the details of the model and the resulting modification to the
matrix $Q$.  It has been shown in \cite{Krajewski98} that the reality
axiom of NCG disallows the inclusion of Majorana fermions, thus excluding
many of the sterile neutrino models which have been studied.  The
inclusion of other fermions in general could also provide a solution,
depending on the representations in which they appear.

\section{Conclusion}
In its NCG setting, the SM, while quite constrained\cite{Brout98}, does
allow for certain extensions which include right-handed neutrinos and
therefore neutrino masses and oscillations.  Our analysis of the
Poincar\'e duality condition has shown that, if we allow only the addition
of right-handed neutrinos, the number of quarks with right-handed states
must be different from the number of neutrinos with right-handed states.

While the option of having a quark without a right-handed state may have
been appealing from the point of view of the strong CP problem, it runs
into trouble with regard to parity violation in the strong interactions.
The case in which two neutrinos are given right-handed states allows for
two independent oscillation lengths, based on our argument that neutrino
oscillations may occur between massive and massless eigenstates.  The
sterile neutrino solution is an interesting one which should be explored
further in the context of NCG.

We should also mention that the almost commutative structure that we have
used here, where the noncommutativity only appears in the finite part of
the algebra, may have to be abandoned at energies near the electroweak
scale.  Then, if all three neutrinos are found to be massive, the
resulting violation of Poincar\'e duality could be viewed as a signal
that the tensor product structure of the spectral triple on which the
current NCG realization of the SM is based is a kind of low-energy limit
of a `truly noncommutative' spectral triple.  In that case the Poincar\'e
duality condition would have to be reevaluated for the spectral triple
from which the one we use today descends.
\par


\begin{references}
\bibitem{Kirsten99}T. A. Kirsten, Rev.\ Mod.\ Phys.\ {\bf 71} (1999) 1213.
\bibitem{Shiozawa99}M. Shiozawa, Nulc.\ Instr.\ and Meth.\ A {\bf 433} 
(1999) 307.
\bibitem{Eitel00}K. Eitel, New J.\ Phys.\ {\bf 2} (2000) 1.
\bibitem{BoehmVogel92}F. Boehm and P. Vogel, Physics of Massive
Neutrinos, second edition (Cambridge University, Cambridge, 1992) p.\ 91.
\bibitem{Iochum96}B. Iochum and T. Sch\"ucker, Commun.\ Math.\ Phys.\   
{\bf 178} (1996) 1.
\bibitem{Gracia-Bondia95}J. M. Gracia-Bond\'{\i}a, Phys.\ Lett.\ B {\bf
351} (1995) 510.
\bibitem{Martin98}C. P. Martin, J. M. Gracia-Bond\'{\i}a, and J. C.
V\'arilly, Phys.\ Rep.\ {\bf 294} (1998) 363.
\bibitem{Connes95}A. Connes, J.\ Math.\ Phys.\ {\bf 36} (1995) 6194.
\bibitem{Connes96}A. Connes, Commun.\ Math.\ Phys.\ {\bf 182} (1996) 155.
\bibitem{Kayser98}B. Kayser, Euro.\ Phys.\ J.\ {\bf C3} (1998) 307.
\bibitem{Kaplan86}D. B. Kaplan and A. V. Manohar, Phys.\ Rev.\ Lett.\ {\bf
56} (1986) 2004.
\bibitem{Donoghue92}J. F. Donoghue and D. Wyler, Phys.\ Rev.\ D {\bf 45} 
(1992) 892.
\bibitem{Banks94}T. Banks, Y. Nir, and N. Seiberg, hep-ph/9403203.
\bibitem{Peccei89}R. D. Peccei, in: CP Violation, ed. C. Jarlskog (World
Scientific, Singapore, 1989) p.\ 503.
\bibitem{Cheng88}H.-Y. Cheng, Phys.\ Rep.\ {\bf 158} (1988) 1.
\bibitem{Zuber98}K. Zuber, Phys.\ Rep.\ {\bf 305} (1998) 295.
\bibitem{Karlen98}D. Karlen, Euro.\ Phys.\ J.\ {\bf C3} (1998) 319.
\bibitem{Caldwell93}D. O. Caldwell and R. N. Mohapatra, Phys.\ Rev.\ D
{\bf 48} (1993) 3259.
\bibitem{Peltoniemi93}J. T. Peltoniemi and J. W. F. Valle, Nucl.\ Phys.\ B
{\bf 406} (1993) 409.
\bibitem{Krajewski98}T. Krajewski, J.\ Geom.\ Phys.\ {\bf 28} (1998) 1.
\bibitem{Brout98}R. Brout, Nucl.\ Phys.\ Proc.\ Suppl.\ {\bf 65} (1998) 3.
\end{references}
\end{document}